\documentstyle[11pt,aaspp4,amssym]{article}      

\newcommand{\lognh}{ \log\,N_H }
\newcommand{\rosat}{{\sl ROSAT}}
\newcommand{\asca}{{\sl ASCA}}
\newcommand{\gis}{{\sl GIS}}
\newcommand{\gistwo}{{\sl GIS2}}
\newcommand{\gisthree}{{\sl GIS3}}
\newcommand{\sis}{{\sl SIS}}
\newcommand{\siszero}{{\sl SIS0}}
\newcommand{\sisone}{{\sl SIS1}}
\newcommand{\pspc}{{\sl PSPC}}
\newcommand{\hri}{{\sl HRI}}

\newcommand{\mdot}{ \dot{M} }
 
\newcommand{\etal}{{\it et al.\ }}
 
\newcommand{\cm}{{\,\rm cm\,}}

\newcommand{\s}{{\,\rm s\,}}

\newcommand{\yr}{{\,\rm yr\,}}
\newcommand{\kms}{{\mbox{$\,\rm km~s^{-1}$}}}
\newcommand{\msun}{\,M_{\sun}\,}

\newcommand{\erg}{{\,\rm erg\,}}

\newcommand{\keV}{{\,\rm keV\,}}

\newcommand{\K}{{\,\rm K\,}}

\begin{document} 

\title{Recent X-Ray Observations of SN1986J with \asca\, and \rosat} 
 
\author{John C. Houck\altaffilmark{1}, Joel N. Bregman\altaffilmark{2}} 
\affil{University of Michigan, 
Department of Astronomy, Dennison Bldg., Ann Arbor, MI, 48109-1090} 
\altaffiltext{1}{MIT, Center for Space Research, 
70 Vassar St., Bldg. 37-662B, Cambridge, MA 02139-4307, houck@space.mit.edu}
\altaffiltext{2}{jbregman@astro.lsa.umich.edu}
\author{Roger A. Chevalier\altaffilmark{3}} 
\affil{University of Virginia, Department of Astronomy, 
P.O. Box 3818, University Station, Charlottesville, VA 22903-0818} 
\altaffiltext{3}{rac5x@virginia.edu}
\author{Kohji Tomisaka\altaffilmark{4}} 
\affil{Laboratory of Physics, Faculty of Education, Niigata
University, Ikarashi 2-8050, Niigata 950-21, Japan} 
\altaffiltext{4}{tomisaka@pulsar.mtk.nao.ac.jp}

\begin{abstract} 
 
We have presented \asca\, and \rosat\, observations of SN 1986J covering
the period 1991 August to 1996 January.  From observations with the
\rosat\, \hri\, and \pspc, we find that the 0.5-2.5 keV flux decreased
$\propto t^{-2}$ during this period; the \asca\, data are consistent
with this result and extend it to the 2-10 keV band.  \asca\, spectra
from 1994 January and 1996 January are consistent with thermal
emission from a solar metallicity plasma at an equilibrium temperature
kT = 5-7.5 keV, somewhat hotter than that observed from other X-ray
supernovae.  These spectra also show a clear Fe K emission line at 6.7
keV with FWHM $< 20,000 \kms$ (90\% confidence). This limit on the
line width is consistent with the reverse shock model of Chevalier \&
Fransson (1994), but does not rule out the clumpy wind model of Chugai
(1993).
 
\end{abstract} 
 
 
\section{Introduction} 
 
A few supernovae have been detected as sources of soft X-ray emission
in the first decades after the supernova event (Canizares Kriss \&
Feigelson 1982; Tanaka 1989; Gorenstein, Hughes \& Tucker 1994; Bregman
\& Pildis 1992; Ryder \etal 1993; Zimmerman, \etal 1993; Tanaka 1993;
Fabian \& Terlevich 1996).  All of the detected sources are associated
with Type II supernovae and, judging from radio observations, the most
X-ray luminous of these appear to have had extensive circumstellar
winds (Schlegel 1995 and references therein).  The interaction between
the shock wave generated by the supernova with a dense circumstellar
environment is expected to produce both soft X-ray emission and
synchrotron radio emission (Chevalier 1982; Chevalier
\& Fransson 1994, hereafter CF).  There are two competing models for the
production of this soft X-ray emission; here we present observations
that help to discriminate between the two models and define the
constraints for further model development.
 
Type II supernovae are associated with massive stars that, during
their supergiant phase, drove slow, dense winds prior to the supernova
event.  If the wind material is smoothly distributed, the exploded
star will drive a fast high temperature shock (the blast shock) into
the wind, with velocities of 10,000-20,000 $\kms$ and temperatures of
$10^9 \K$.  Also, a relatively slow shock (the reverse shock)
propagates back into the stellar envelope at a velocity of 500-1000
$\kms$ relative to the expanding ejecta.  This relatively
mild shock-deceleration of the stellar envelope produces $10^7 \K$ gas
that can account for the soft X-ray emission detected with X-ray
telescopes.  Little X-ray radiation emerges from the dilute $10^9 \K$
gas behind the blast shock, although particle acceleration in the
shocked shell can produce non-thermal radio emission (Chevalier 1982; CF).
 
The situation can be substantially different if either the 
circumstellar wind material or the exploded stellar envelope is highly
clumped.  The case of a highly clumped circumstellar wind has been 
studied by Chugai (1993) and applied to SN 1986J.  His study was 
motivated by the observation of narrow Balmer emission lines ($<600 
\kms$; Leibundgut et al. 1991) in the optical spectrum of SN 1986J. 
This observation was surprising since hydrogen emission lines are 
usually identified with the outer part of the exploded star, which for
SN 1986J, is expanding at $\sim$10,000 $\kms$ (based upon its age and radio 
VLBI size; Bartel, Rupen, and Weiler 1989). Chugai (1993) proposed 
that the progenitor wind was clumpy and that the narrow emission lines
and the soft X-ray emission originate in the wind clumps or clouds. 
In this model, the reverse shock is too weak to produce observable 
soft X-rays.  Rather, it is the pressure in the strongly shocked wind 
that crushes the embedded clouds, driving shocks into the clouds.  The
observed soft X-rays are produced by radiative cooling of shocked 
cloud material, and the Balmer emission is produced when these soft 
X-rays are reprocessed in the cooler cloud material.  The inertia of 
the clouds is so great that they are not accelerated outward when the 
blast shock passes over them, so the FWHM of the optical emission 
lines reflects only the speed of the shock driven into the clouds. 
Because the reverse shock model predicts X-ray lines of order 10,000 
$\kms$ wide, the line widths of the X-ray emission lines, which can be
observed with \asca, can be a powerful discriminator between the 
reverse shock model and model of Chugai (1993). 
 
X-ray emission in young supernovae has been observed in only a few 
cases (SN 1978K, 1980K, 1986J, 1987A, 1988Z, 1993J, and 1995N), and of these,
only two are bright enough to obtain both spectra and fluxes: SN 1986J
and 1993J.  SN1986J, in the nearby edge-on spiral galaxy NGC 891, was 
first discovered in the radio (Rupen et al. 1987) and is one of the 
closest recent supernovae.  Because of its large radio luminosity, 
SN1986J is thought to have had a progenitor red supergiant star with 
an unusually dense wind (Weiler, Panagia, and Sramek 1990) and was 
predicted to be luminous in soft X-rays (Chevalier 1987).  An August 
1991 \pspc\, observation (Bregman and Pildis 1992) confirmed 
that it is X-ray luminous.  Based on this \pspc\, observation, the X-ray
emission was characterized by $L_x (0.1-2.5 \keV) = (1.6-7) \times 
10^{40} \erg\s^{-1}$, $T_x= 1.0-3.9 \keV$, and an absorbing column of 
$(5-14) \times 10^{21}~\cm^{-2}$.  Because of the high absorption column,
most of the photons were removed from the low energy band of the
\rosat\ \pspc\, so that the received X-ray photons fell into a very limited
energy range (0.7-2 keV), and the temperature and luminosity were
poorly determined.  Nevertheless, these \pspc\, measurements, when
combined with radio data, gave a consistent fit to the reverse shock
model (CF).
 
Four new observations were obtained during the past five years: a 
second \rosat\, \pspc\, observation, a \rosat\, \hri\, observation, and two 
\asca\, observations.  The \rosat\, and \asca\, observations provide clear
evidence of dimming while the \asca\, observations determine the
abundance, temperature, and line width of the X-ray emitting material.
Here, we present these recent observations and compare these new
results with the theoretical expectations for the competing models.
 
\section{Observations and Analysis} 
\label{obs} 
 
\subsection{\rosat\, Observations of SN 1986J} 
 
\subsubsection{\pspc\, Observations} 
 
The first observation of SN 1986J, obtained on August 18, 1991 (as
part of AO1), consists of 24.9 ksec of live time data, which included
some periods where the background was more than twice the modal
background level; nearly all of the higher background levels are
identified with scattered solar X-rays.  We discarded observing
periods when the background exceeded 1.7 times the modal background
value, which yielded a remaining live time of 17.1 ksec.  In the
initial analysis of these data (Bregman and Pildis 1992), no data were
discarded, but the difference in spectral fits and fluxes between the
two types of analysis are insignificant.  The second observation of SN
1986J was obtained 23 months later, during the period July 8-27, 1993
(as part of AO4), and consists of 31.5 ksec of useful observing time.
There are no periods during which the background became high relative
to the modal value of the background, so no data were excluded.  The
background, as well as being better behaved than the 1991
observations, has a lower mean value, so low intensity features, such
as those associated with the diffuse galactic emission, are seen at a
higher signal-to-noise level (Bregman \& Houck 1996).

\begin{figure}
%\plottwo{f1a.eps}{f1b.eps}
\plotfiddle{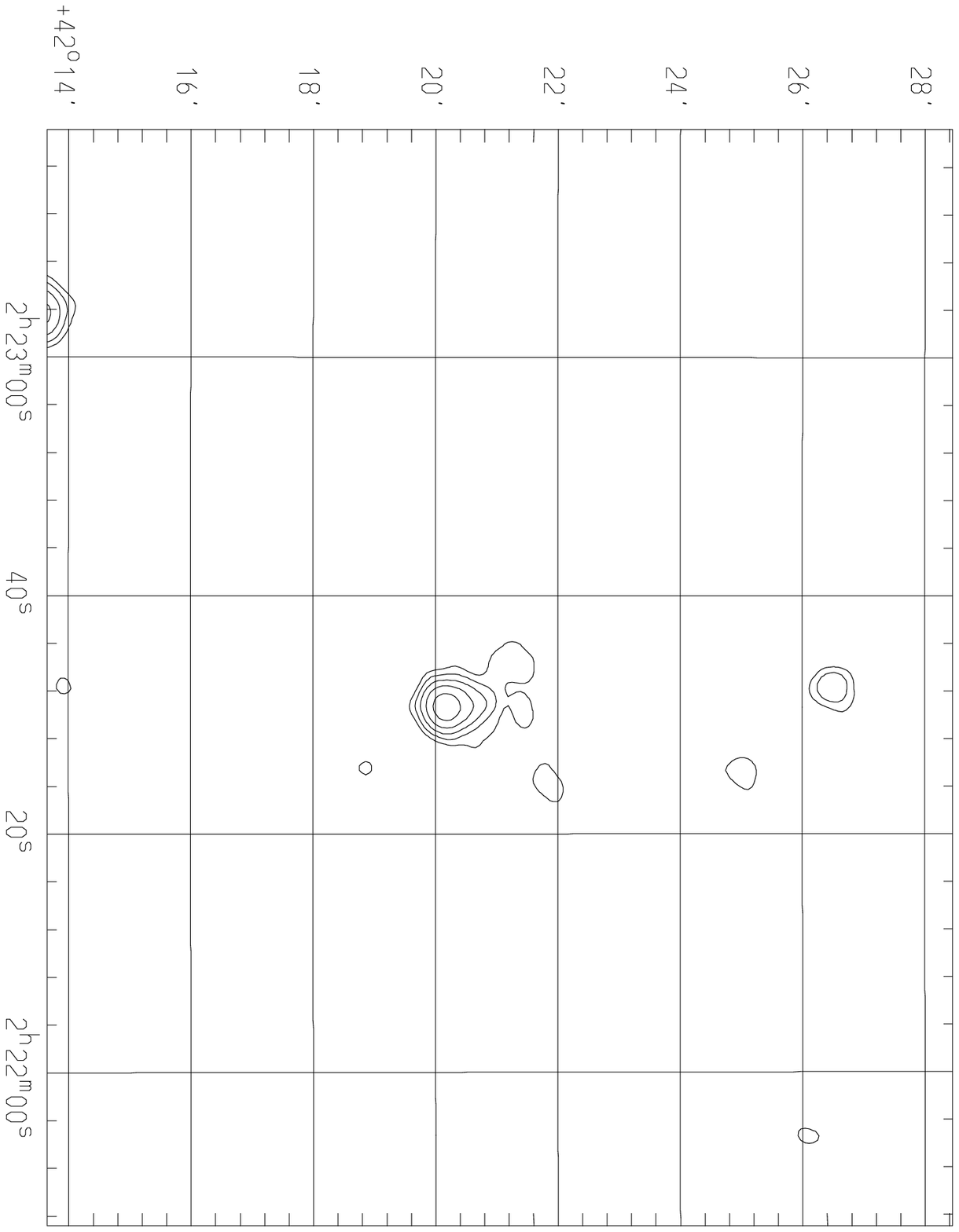}{3in}{90}{55}{55}{230}{-20}
\plotfiddle{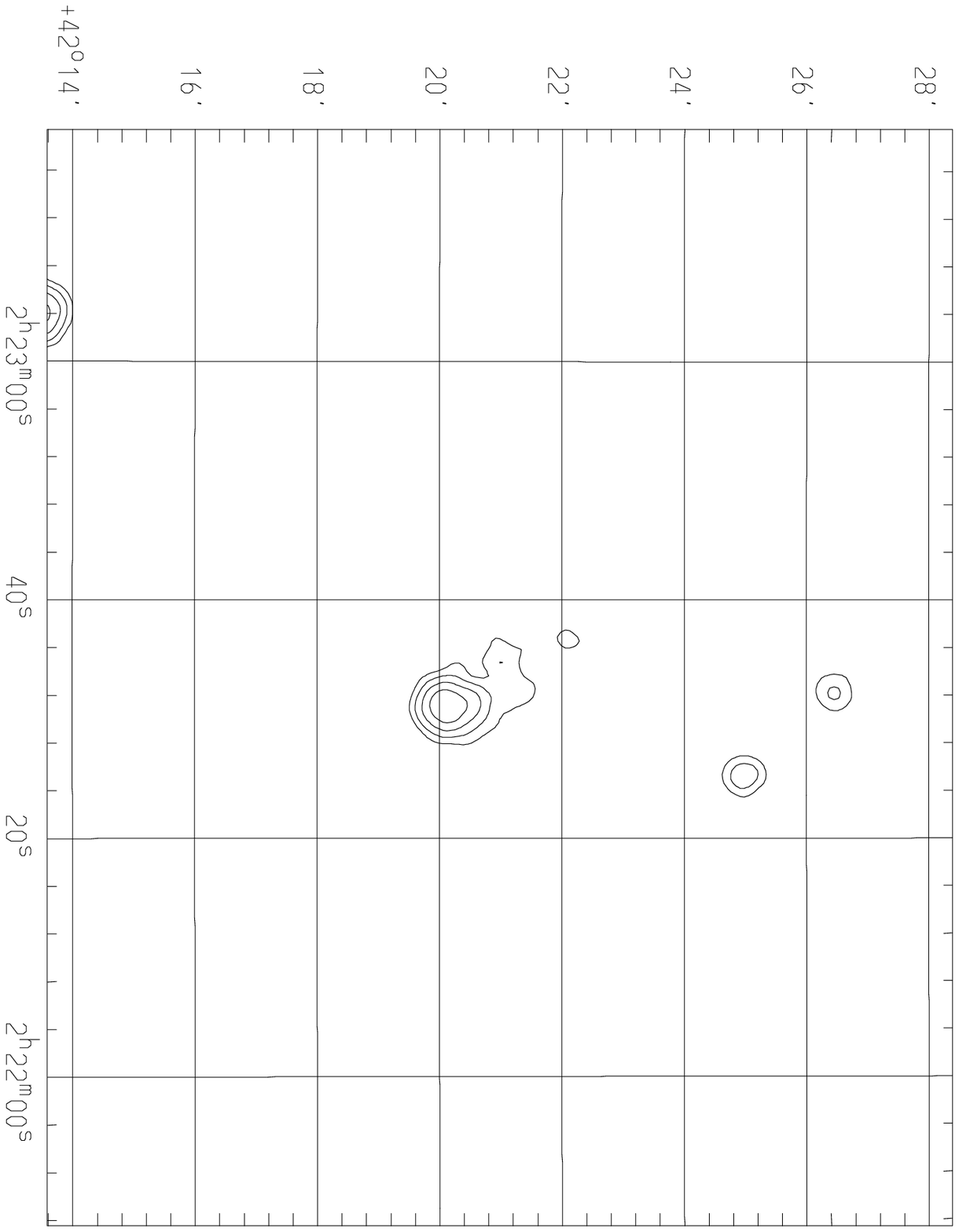}{3in}{90}{55}{55}{230}{-30}
\caption{\pspc\, images of SN1986J from a)
1991 Aug and b) 1993 Jul.  
In (a) the contours represent 0.3, 0.6, 1.2, 2.4, 
and 4.8 counts per 4\arcsec\, square pixel.  In (b) the contour
levels have been
scaled to the same effective exposure time as in (a), which means that
the contour levels are 0.55, 1.11, 2.22, 4.43, and 8.86 counts per 
4\arcsec\, square pixel.
The supernova image is elongated because of the weaker source ``Xnorth''
27.7\arcsec\, north and 0.4\arcsec\, east of the supernova position.  The
sources are unresolved in the \pspc, but are clearly resolved by the
\hri. Note that most of the extra emission in the field of view is not
noise, but is diffuse emission from NGC 891. 
\label{pspc-images}
} 
\end{figure}

\begin{figure}
\plotfiddle{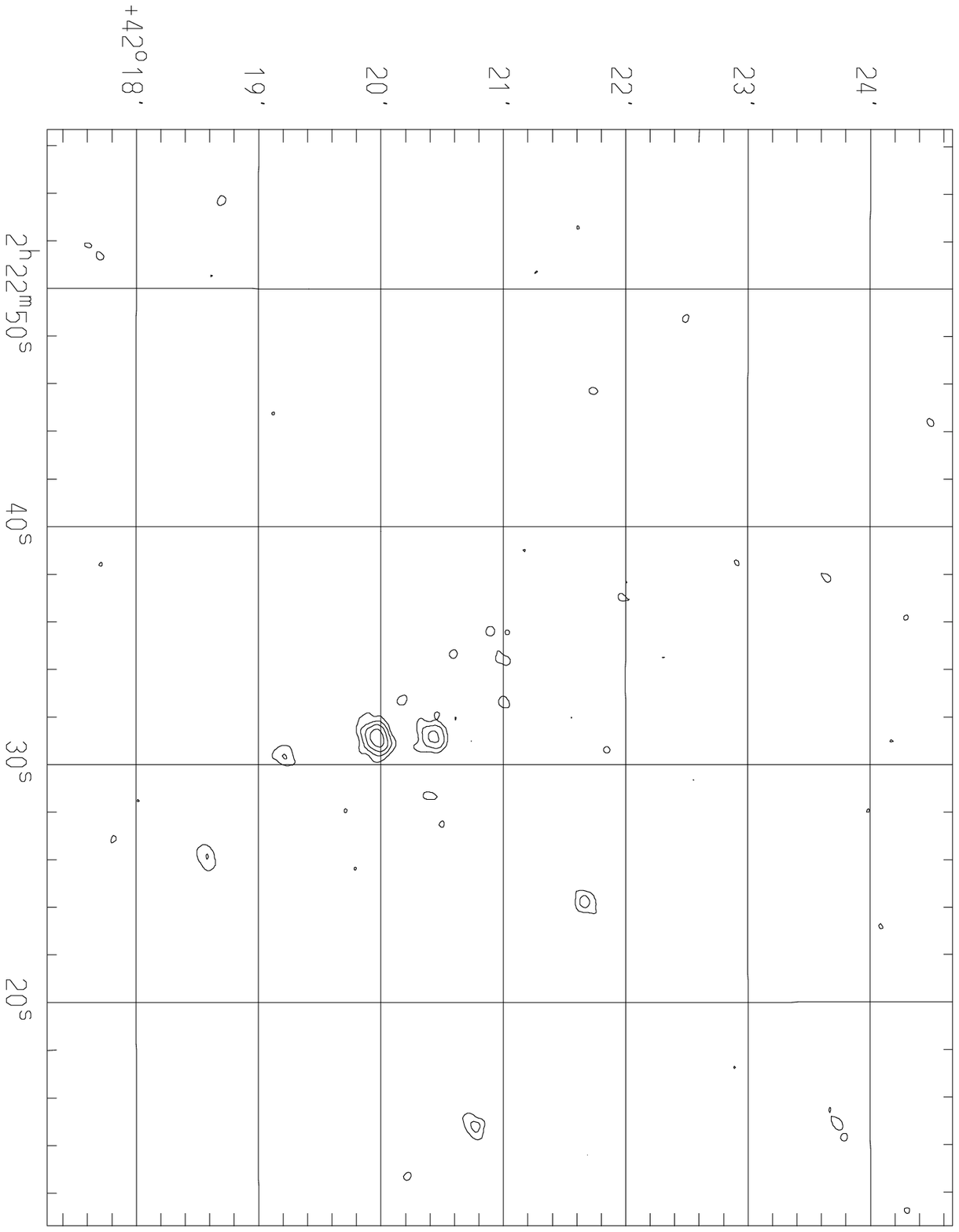}{3in}{90}{60}{60}{260}{-30}
\caption{1995 Jan \hri\, image of SN1986J.  The nearby
weaker point source 
``Xnorth'' is clearly resolved.  The contour levels represent 
0.35, 0.70, 1.4, and 2.8 counts per 1\arcsec\, square pixel. 
Note that most of the extra emission in the field of view is not
noise, but is diffuse emission from NGC 891. 
\label{hri-image}
}
\end{figure}
 
At the resolution of the \pspc, the emission from SN 1986J is
contaminated by a weaker nearby source and by the diffuse galactic
emission.  This weaker source is visible in the \pspc\, image in that it
causes a distortion in the shape of the emission from SN 1986J
(Fig. \ref{pspc-images}a, \ref{pspc-images}b), but in the \hri\, image
(Fig. \ref{hri-image}), the two sources are clearly separated, with
the weaker source 27.7\arcsec\, to the north and 0.4\arcsec\, to the east
(this source will be referred to as Xnorth in the tables).  In the
extraction of \pspc\, fluxes for SN 1986J, we fit point sources to the
data, where the point spread function (PSF) is approximated by a
Gaussian profile and the relative source separation is given by the
\hri\, positions.  The amplitude of the two Gaussians was varied until
the remaining residuals were similar to those expected from photon
noise and where the emission from the underlying galaxy was
approximately continuous (the galaxy contribution is small).  The
resulting count rates for SN 1986J and source Xnorth are listed in
Table \ref{rosat-table}.

\begin{deluxetable}{llllccc}
\footnotesize
\tablewidth{6.5in}
\tablecaption{\rosat\ Observations \label{rosat-table} }
\tablehead{ 
\colhead{Obs.} &
\colhead{MJD} &
\colhead{Instr.} &
\colhead{Source} &
\colhead{Count Rate} &
\colhead{Flux\tablenotemark{a}} &
\colhead{L\tablenotemark{a}} \nl
\colhead{ } &
\colhead{ } &
\colhead{ } &
\colhead{ } &
\colhead{$(10^{-2}~{\rm counts}~{\rm s}^{-1})$} &
\colhead{$({\rm erg}~{\rm s}^{-1}~{\rm cm}^{-2})$} &
\colhead{$({\rm erg}~{\rm s}^{-1})$} 
          }
\startdata
Aug 1991 & 48486.1 & PSPC & SN1986J & $2.38\pm 0.12$ & 7.90(-13) & 9.47(39) \nl
         &         &      & XNorth  & $0.39\pm 0.05$ &           &          \nl
Jul 1993 & 49206.5 & PSPC & SN1986J & $1.44\pm 0.07$ & 4.91(-13) & 5.89(39) \nl
         &         &      & XNorth  & $0.37\pm 0.04$ &           &          \nl
Jan 1995 & 49744.0 &  HRI & SN1986J & $0.43\pm 0.03$ & 3.95(-13) & 4.73(39) \nl
         &         &      & XNorth  & $0.17\pm 0.02$ &           & 
\enddata
\tablenotetext{a}{log$N_H$ = 21.7; kT = 2 keV; energy range 0.5-2.5 keV}
\end{deluxetable}

%-----------------------------------------------------------------------

\begin{figure}
\plotfiddle{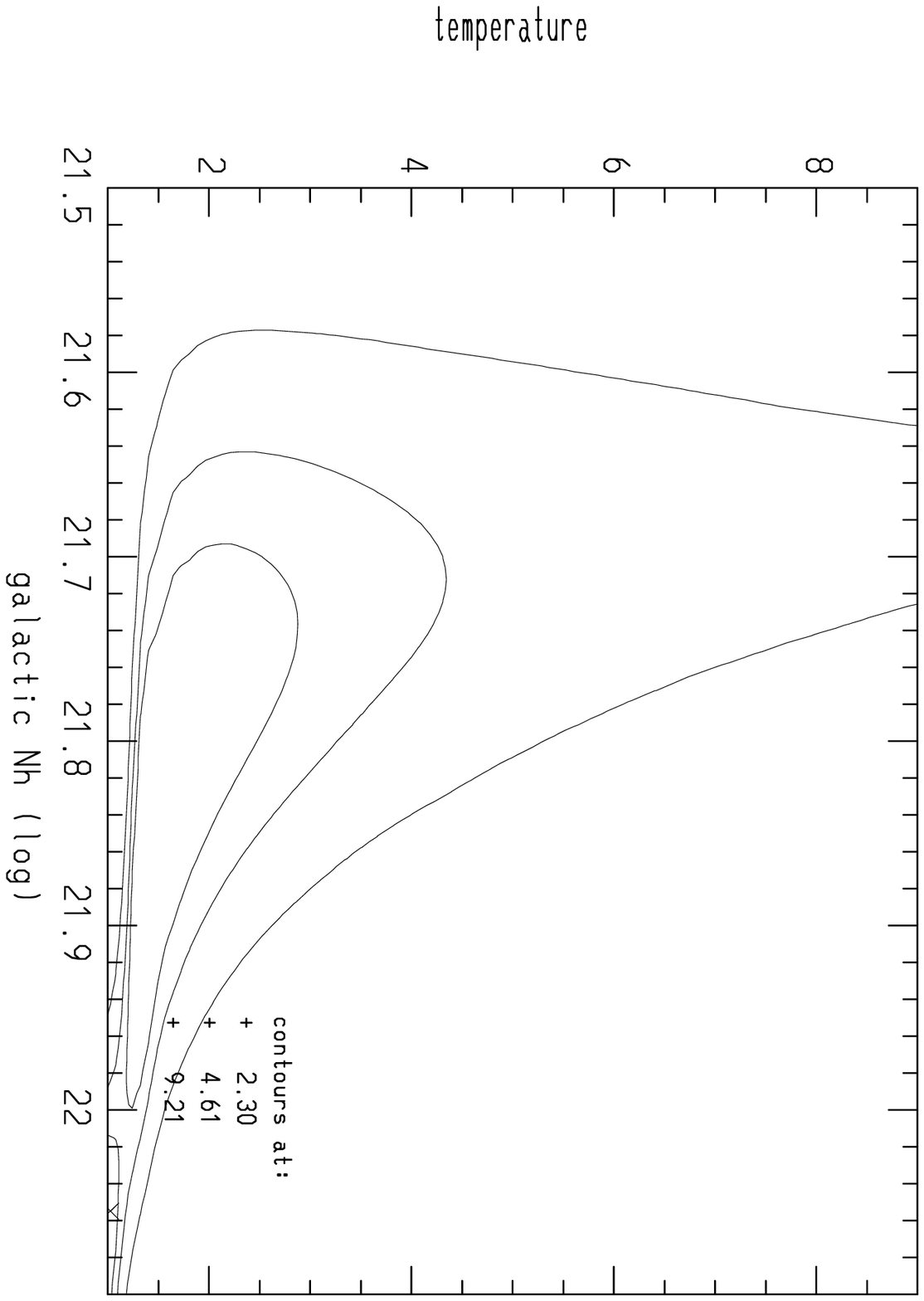}{3in}{90}{55}{55}{230}{-20}
\plotfiddle{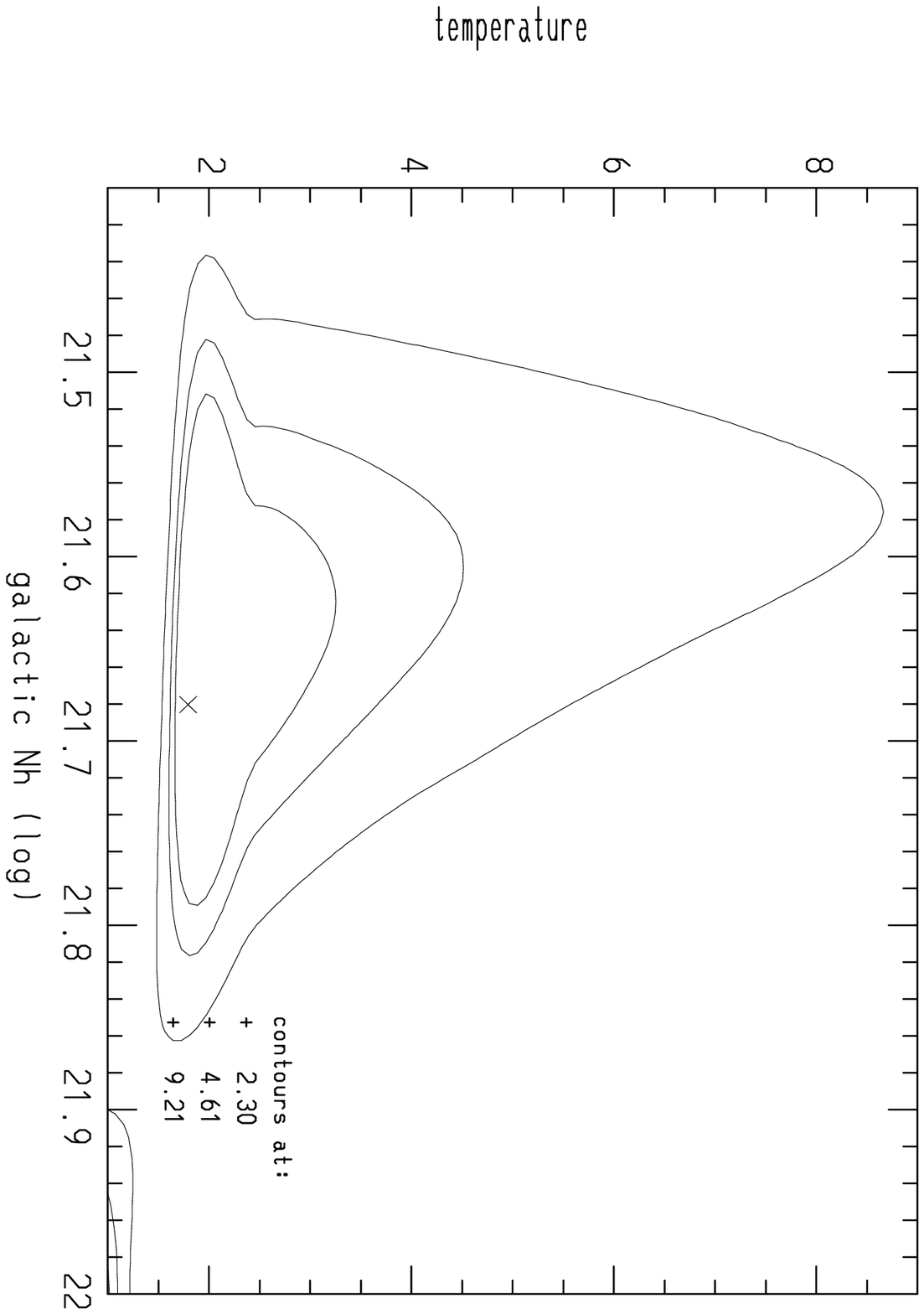}{3in}{90}{55}{55}{230}{-30}
\caption{Temperature and column density
confidence contours for \rosat\ \pspc\, spectra from a) 1991 Aug and
b) 1993 Jul. 
\label{conf-contours}
} 
\end{figure}

For spectral analysis we used a 25\arcsec\, diameter aperture centered
on the supernova minus the background in an annulus of interior and
exterior radii 75\arcsec\, and 200\arcsec.  This background includes the
diffuse emission from the galaxy, which should improve the accuracy of
the resultant spectra, although the number of background photons is not
large in the 25\arcsec\, diameter aperture.  The 25\arcsec\, aperture used
encircles 91\% of photons from the supernova (at 1 keV), and the use
of larger apertures lowered the S/N of the spectrum (the background
was increased) without changing the results of the spectral analysis.
A single-temperature thermal plasma model (a ``Raymond-Smith'' plasma)
provides a successful fit to the data.  Based solely on the \pspc\,
data, there are two regions of temperature-$N_H$ space that produce
equally acceptable spectral fits, a low temperature (0.2-1 keV) high
$N_H$ region (log $N_H$ = 22.0-22.5) and a higher temperature (1-10
keV) lower $N_H$ region (log$N_H$ = 21.5-21.9).  This degeneracy is
broken by the \asca data, which reveal that the lower column, higher
temperature solution is the correct one, and this is the one that will
be discussed.  The pulse-height spectra from the two periods are quite
similar and there was no statistically significant change in the
acceptable spectral fits.  However, there is an indication that $N_H$
has decreased between 1991 and 1993, as seen in the $\chi^2$ grids
(Fig. \ref{conf-contours}a and Fig. \ref{conf-contours}b).  For the sake of comparison, we adopted
two spectral models that give good fits to the two spectra in order to
make comparisons of flux and luminosity.  The two fits have cosmic
abundance and log$N_H$ = 21.7, T = 2.0 keV, which is within the 1\%
contour for each model (Table \ref{rosat-table}), and log$N_H$ =
21.65, T = 5.0 keV, which is about at the 95\% confidence contour, and
represents a higher temperature case. However, the difference in the
absorption corrected flux and the luminosity (columns 6 and 7 in Table
\ref{rosat-table}) were insignificant so only the results of the 2 keV
model are shown in Table \ref{rosat-table}.

\subsubsection{The \hri\, Observation} 
 
The \hri\, observations were obtained during 26 Jan 1995 to 4 Feb 1995
for a total useful exposure time of 97.17 ksec.  The background was
well-behaved during the observation and the image clearly shows the
supernova and the source to the north (Fig. \ref{hri-image}).  Other
lower-level emission due to the galaxy can be extracted from the data
and are discussed in Bregman \& Houck (1996).  The count rate was
determined within a circle of 13\arcsec\, radius, which encircles 90\%
of the photons from a point source; a 10\% correction was applied to
compensate for the scattered photons.  For the determination of the
flux and luminosity, we adopted the same spectrum used for \pspc\,
spectral analysis (log$N_H$ = 21.7, T = 2.0 keV).
 
\subsection{Time Variability} 
 
\begin{figure}
\plotfiddle{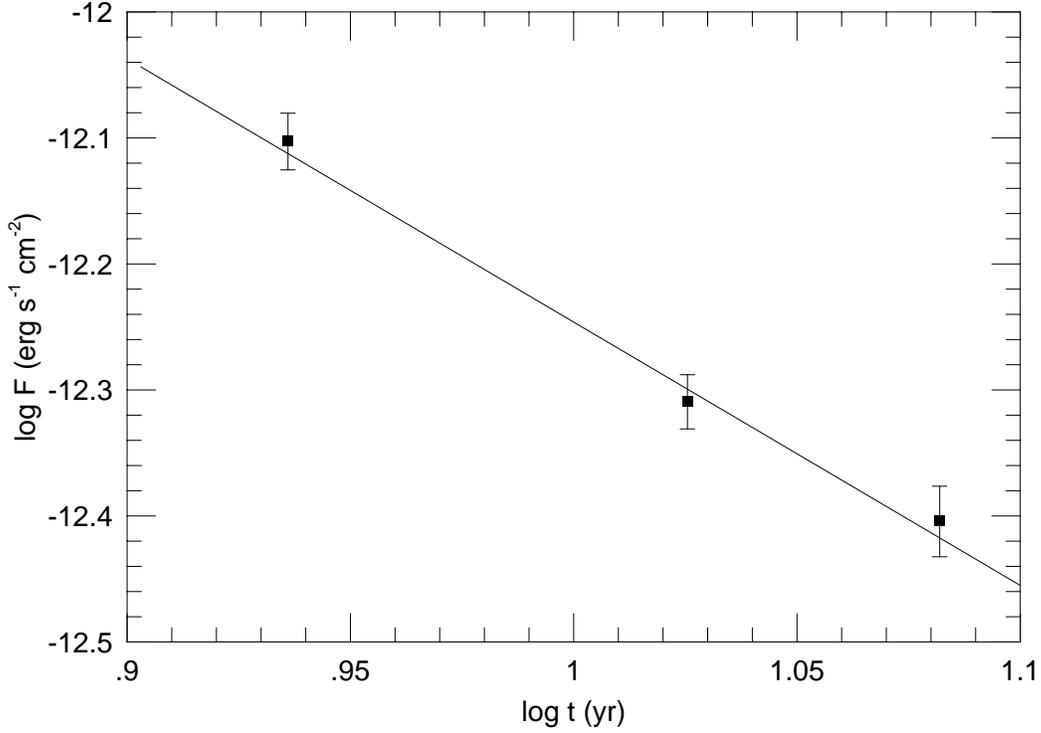}{4in}{0}{55}{55}{-215}{0}
\caption{X-ray light curve of SN1986J from \rosat\ \pspc\
 and \hri\ observations. t is the age in years since 1983.0 and F is the
unabsorbed flux in $\erg \s^{-1} \cm^{-2}$ for an absorbing column of
log $N_H$ = 21.7.
\label{light-curve}
} 
\end{figure}

The clearest difference between the observations is that there has 
been a significant decrease in the count rate, flux, and luminosity 
over a three and a half year period.  The supernova event probably 
occurred in the last half of 1982 or the first few months of 1983, so 
if we assume that the event occurred in 1983.0, then the first \pspc\, 
observation occurred 8.63 years after the SN event, the second \pspc\, 
observation occurred 10.61 after the SN and the last \rosat\, 
observation was obtained when the remnant was 12.08 years old.  The 
time variation is decreasing approximately as $t^{-2}$, based upon 
these three measurements (Fig. \ref{light-curve}), although with only three 
points, the form of the dimming is not well-established. 

\subsection{\asca\, Observations of SN 1986J} 
 
\begin{figure}
\plotfiddle{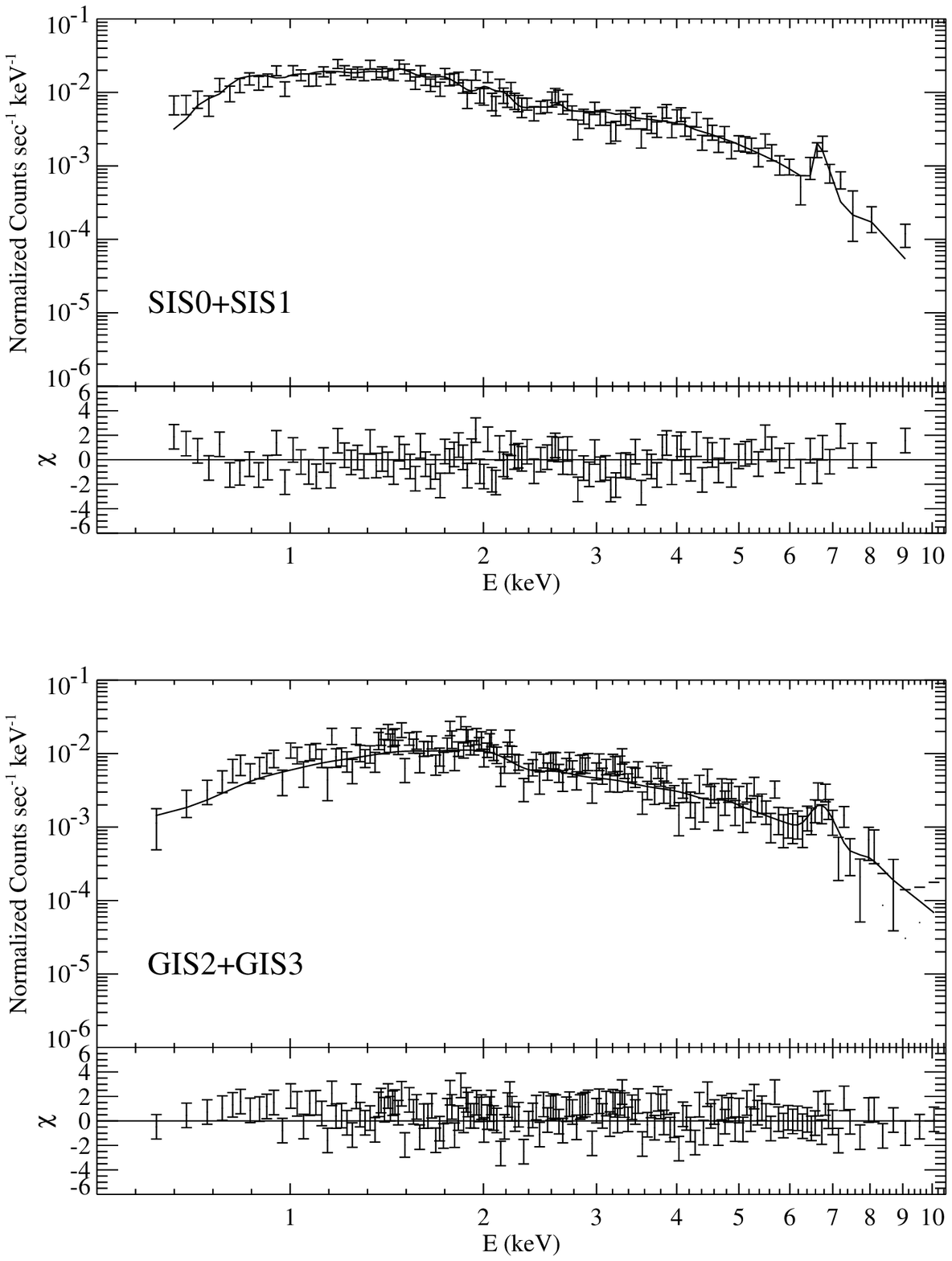}{6.75in}{0}{90}{90}{-200}{-20}
\caption{Sum of a) \siszero\ and \sisone\ and b) \gistwo\ and
\gisthree\ spectra from 1994 Jan along with the $\chi^2$ values 
for model 2a94; the solid line is the model fit, after being
folded through the instrumental response.
\label{jan94-spectra}
}
\end{figure}

\begin{figure}
\plotfiddle{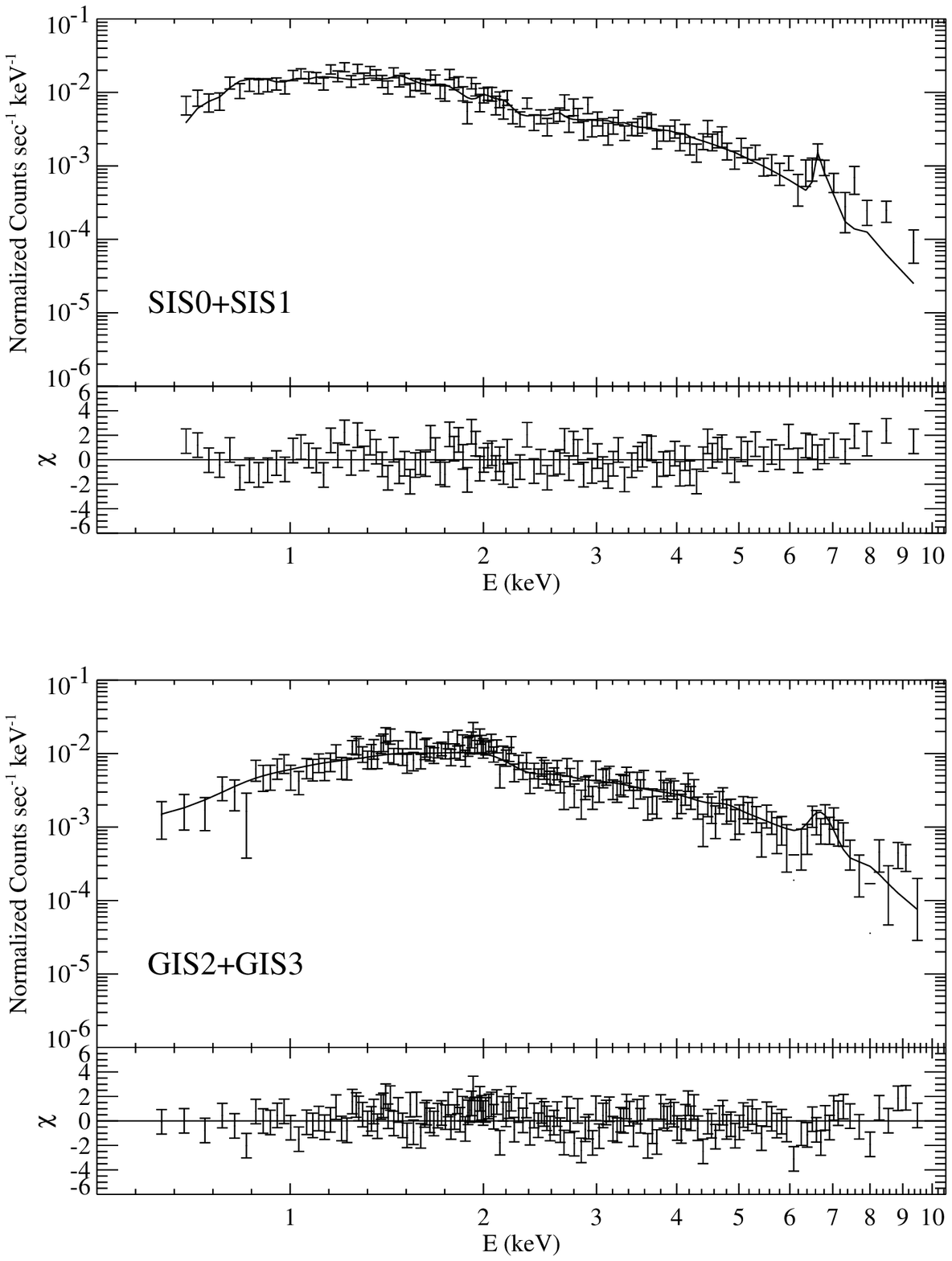}{6.75in}{0}{90}{90}{-200}{-20}
\caption{Sum of the a) \siszero\ and \sisone\ and b)
\gistwo\ and \gisthree\ spectra from 1996 Jan along with the $\chi^2$ values
for the model 2a96; the solid line is the model fit, after being
folded through the instrumental response.
\label{jan96-spectra}
}
\end{figure}

SN1986J was observed by \asca\, during 1994 January 21-22 and again
during 1996 January 30-31.  Standard criteria were used to exclude
data taken during high noise conditions, especially those taken in
close proximity to the South Atlantic Anomaly and other regions of low
geomagnetic rigidity, and those taken when the line of sight to the
supernova lay too close to the limb of the earth.  We excluded \gis\ 
data taken while the ambient magnetic rigidity was less than 6 GeV/c
and when the elevation angle above the limb of the earth was less than
5 degrees.  For the \sis\, instruments, the cutoff values were 6 GeV/c
and 10 degrees (20 degrees for the bright earth), respectively.
Charged particle events were excluded from the \sis\, data by retaining
only \sis\, events with grade 0,2,3 and 4.  X-ray detections were
identified and extracted using GISCLEAN and SISCLEAN within the
XSELECT reduction package.  The net usable exposure times along with
the count rates for the extraction regions used (radius $R_{\rm extr}$
arcmin) are given in Table \ref{asca-table-1}. The \gis\ extraction
regions were chosen to minimize contamination from a transient X-ray
source about 6\arcmin\, NE of the SN position along the galactic
disk. Background spectra were obtained from archival blank sky
observations using the same extraction regions used for the source
spectra.  The background subtracted spectra are shown in Figs
\ref{jan94-spectra} and \ref{jan96-spectra}.

%-----------------------------------------------------------------------

%\input asca_table

% ensures one table per page
%\clearpage

\begin{deluxetable}{lccccccccc}
\tablewidth{6.5in}
\tablecaption{\asca\ Observations \label{asca-table-1}}
\tablehead{ 
\colhead{ } &
\colhead{ } &
\multicolumn{4}{c}{MJD 49374.5 (1994 Jan)} &
\multicolumn{4}{c}{MJD 50113.5 (1996 Jan)\hfil} \nl
\colhead{Instr.} &
\colhead{$R_{\rm extr}$} &
\colhead{Exp.} &
\multicolumn{3}{c}{($10^{-2}~{\rm counts~s}^{-1}$)} &
\colhead{Exp.} &
\multicolumn{3}{c}{($10^{-2}~{\rm counts~s}^{-1}$)} \nl
\colhead{ } &
\colhead{(arcmin)} &
\colhead{(ksec)} &
\colhead{total} &
\colhead{bkgd} &
\colhead{src} &
\colhead{(ksec)} &
\colhead{total} &
\colhead{bkgd} &
\colhead{src} 
          }
\startdata
SIS0 & 2.97 &  43.8 & 5.11 & 1.15 & 3.96 &   50.0 & 4.46 & 1.19 & 3.27 \nl
SIS1 & 2.97 &  41.4 & 4.48 & 1.07 & 3.41 &   49.8 & 4.13 & 1.27 & 2.86 \nl
GIS2 & 4.70 &  47.1 & 4.24 & 1.34 & 2.90 &   54.4 & 3.59 & 1.25 & 2.34 \nl
GIS3 & 4.70 &  47.2 & 5.20 & 1.56 & 3.64 &   54.4 & 4.48 & 1.48 & 3.00 \nl    
\enddata
\end{deluxetable}

%-----------------------------------------------------------------------

X-ray emission from SN 1986J is superposed on the weak, extended X-ray
emission from the host galaxy.  \rosat\, \pspc\, observations of NGC 891
detect soft X-rays extending to 1\farcm\,7 above the plane of the
galaxy and 2\farcm\,2 radially along the disk of the galaxy
(Bregman \& Pildis 1994; Bregman \& Houck 1996).
Because of the broad ($\sim3\arcmin$ half-power radius), asymmetric
and energy dependent point spread function (PSF) of the \asca\, X-ray
telescope (XRT), X-ray emission from the supernova is confused with
X-ray emission from the galaxy and the nearby point source seen with
the \rosat\ \hri.  This makes it difficult to determine the total X-ray
flux from the supernova and confuses the interpretation of the X-ray
spectrum at low energies, $\lesssim$ 1 keV, where emission from the host
galaxy may dominate (Bregman \& Pildis 1994).
 
For each epoch, we simultaneously fit the observed spectra from all
four instruments with one and two component thermal emission models
using the MEKAL model in XSPEC.  The important model parameters are
the plasma temperature ($kT$) and metal abundance ($Z$), the absorbing
column density ($N_H$) and the normalization $K = (10^{-14}/4\pi
D^2)\int n_e^2 dV$ where $D$ is the distance to the source (cm) and
$n_e$ is the electron density ($\cm^{-3}$).  For consistency with the
$\chi^2$ minimization procedure, all spectra were binned to contain at
least 20 counts per bin.  The best fit parameters and 90\% confidence
limits for selected parameters are given in Table
\ref{model-table}.

\begin{deluxetable}{rlllllll}
\footnotesize
\tablewidth{6.5in}
\tablecaption{\asca\ Spectrum Models \label{model-table}}
\tablehead{
  \colhead{Model} &
 \colhead{$\chi^2/N$} &
 \multicolumn{4}{c}{High Temperature Component\tablenotemark{a}} &
 \multicolumn{2}{c}{\hspace{-6mm}Model Flux\tablenotemark{b}} \nl
 \colhead{ } &
 \colhead{ } &
 \colhead{$kT_1$} &
 \colhead{$Z_1$\tablenotemark{c}} &
 \colhead{$N_{H,1}$\tablenotemark{d}} &
 \colhead{$K_1$\tablenotemark{e}} &
 \colhead{$F_{0.5-2.5}$} &
 \colhead{$F_{2-10}$} \nl
 \colhead{ } &     % under model
 \colhead{ } &     % under chi^2/N
 \colhead{(keV)} & % under kT_1
 \colhead{ } &     % under Z_1
 \colhead{ } &     % under NH_1
  \colhead{ } &    % under K_1
 \colhead{ } &     % under F(0.5-2.5 keV)
 \colhead{ }       % under F(2-10 keV)
          }
\startdata
2a94 & 360.9/334 
            & 6.63(5.99-7.39) & 1.06(0.78-1.38) & 0.5             & 1.31(-3)
            & 0.606 & 1.63 \nl
2ac94 & 395.4/334
            & 5.66(5.18-6.21) & 0.98(0.74-1.26) & 0.5             & 1.39(-3)
            & 0.597 & 1.57 \nl
2af94 & 361.0/334
            & 6.62(5.98-7.37) & 1.0             & 0.5             & 1.32(-3)
            & 0.606 & 1.61 \nl
2b94 & 394.0/334
            & 10.3            & 1.63            & 0.275           & 1.07(-3)
            & 0.633 & 1.70 \nl
2c94 & 380.6/334
            & 4.82            & 0.99            & 0.725           & 1.52(-3)
            & 0.590 & 1.54 \nl
1t94 & 372.4/334
            & 9.24(8.16-11.3) & 1.36(0.98-1.96) & 0.20(0.15-0.23) & 1.11(-3)
            & 0.611 & 1.67 \nl
     & & & & &  & & \nl
2a96 & 328.0/314
            & 5.37(4.86-5.99) & 0.64(0.42-0.90) & 0.5             & 1.18(-3)
            & 0.528 & 1.17 \nl
2ac96 & 348.9/314
            & 4.50(4.14-4.92) & 0.66(0.45-0.89) & 0.5             & 1.27(-3)
            & 0.520 & 1.12 \nl
2af96 & 333.2/314
            & 5.36(4.88-5.95) & 1.0             & 0.5             & 1.09(-3)
            & 0.524 & 1.20 \nl
2b96 & 354.5/314
            & 8.55            & 0.81            & 0.275           & 9.69(-4)
            & 0.547 & 1.25 \nl
2c96 & 344.3/314
            & 3.99            & 0.72            & 0.725           & 1.37(-3)
            & 0.515 & 1.11 \nl
1t96 & 326.2/314 
            & 7.84(6.72-9.51) & 0.72(0.43-1.07) & 0.16(0.12-0.20) & 0.99(-3)
            & 0.532 & 1.21
\enddata
\tablenotetext{a}{For the two temperature models, the low temperature
component was fixed with $K_2=5.3\times 10^{-5}$, $Z_2 = 1$, $\log N_H = 21.0$.
Models 2a9X, 2af9X, 2b9X and 2c9X had $kT_2 = 0.62 \keV$, while models 2ac9X
had $kT_2 = 0.26~\keV$. }
\tablenotetext{b}{detected flux in units $10^{-12}~\erg\s^{-1}\cm^{-2}$}
\tablenotetext{c}{Fe abundance relative to solar}
\tablenotetext{d}{absorbing column in units $10^{22}~\cm^{-2}$}
\tablenotetext{e}{spectrum normalization coefficient $K_i \equiv (10^{-14}/4\pi
D^2)\int n_e^2 dV$ where $D$ is the distance to the source (cm)
and $n_e$ is the electron density ($\cm^{-3}$)}
\end{deluxetable}

%-----------------------------------------------------------------------
 
Models 1t94 and 1t96 are single temperature fits to the 1994 Jan and
1996 Jan observations respectively. These fits are about the same
within the errors, indicating that the supernova spectrum did not
change drastically over this two year period.  Closer inspection shows
that the single temperature fits 1t94 and 1t96 yield significantly
higher temperatures $kT \gtrsim 6.5 \keV$ than either the $kT=1.0-3.9
\keV$ deduced from earlier \rosat\, observations of SN1986J
Bregman \& Pildis (1992) or the $kT \sim 3.0
\keV$ deduced for SN1978K Petre \etal (1994).   
Also, the 90\% confidence limits on the absorbing column density
($\log N_H= 21.18-21.42$) are significantly lower than the
\pspc\, limits on the absorbing column and the absorbing column deduced
from other observations.

From 21 cm observations, Heiles (1975), finds that the
\ion{H}{1} column density within our galaxy along this line of sight 
is $N($\ion{H}{1} $) = 7.3 \times 10^{20} \cm^{-2}$. Rupen
\etal (1987) used VLA observations at 21 cm to measure a column
density of $N($\ion{H}{1}$) = 2 \times 10^{21} \cm^{-2}$ within NGC
891 along our line of sight toward SN1986J.  Therefore, the
total \ion{H}{1} column along the line of sight to the supernova is
$N($\ion{H}{1}$) = 2.7 \times 10^{21} \cm^{-2}$, corresponding to
an optical extinction $A_V = 1.5$, consistent with that used by
Leibundgut \etal (1991).  The X-ray absorbing column 
includes both the neutral and the warm ionized components along the
line of sight.  By comparing the mean vertical distributions of
\ion{H}{1} and \ion{H}{2} at the solar circle (Dickey \& 
Lockman 1991;Reynolds 1991), we estimate that the 
\ion{H}{1} column density is about 75\% of the total 
\ion{H}{1} + \ion{H}{2} column.  Therefore, the total X-ray absorbing
column density should be about $\lognh = 21.56$, about 40\% larger
than the 90\% confidence upper limit from our single temperature model
fits.
 
The lack of agreement between the \asca\, absorbing column and the other
observations is unsurprising; because of the lack of sensitivity $\lesssim
0.8 \keV$, we do not expect the \asca\, observations to strongly
constrain the absorbing column.  Furthermore, because the \asca
spectra contain X-ray emission from at least three components
(SN1986J, diffuse emission from NGC 891, and the nearby point source
in the disk), a single temperature spectrum model is an obvious
oversimplification.  In the following, we examine a range of
two-temperature models in which the absorbing column associated with
the stronger (high temperature) component is constrained to the range
$N_H = (2.75-7.25)\times 10^{21} \cm^{-2}$, consistent with our
\rosat\ \pspc\, spectra. Models 2a94, 2b94 and 2c94 are two-temperature
fits to the 1994 Jan \asca\, spectra in which the absorbing column
associated with the high temperature component was fixed at three
different values within the range allowed by the Jan 1993 \rosat\ \pspc\,
observation.  Models 2a96, 2b96 and 2c96 are the corresponding
sequence of two-temperature fits to the 1996 Jan
\asca\, spectra. In all of the two-temperature fits, the parameters
associated with the low temperature component were fixed; in models
2a9X, 2b9X and 2c9X, the temperature and normalization values were
chosen by fitting models to the individual \sis\ instruments at both
epochs.  The absorbing column for the low temperature component was
fixed at $\log N_H = 21.0$, consistent with Galactic absorption along
the line of sight to NGC891.  Model 2ac9X is the same as model 2a9X
except $kT_2=0.26~\keV$, consistent with the PSPC observations of the
diffuse emission around NGC891 (Bregman
\& Pildis 1994; Bregman \& Houck 1996). Model 2af9X is the
same as model 2a9X except that the metallicity is fixed at solar. 
 
Models 2a94 and 2a96 (and 2ac9X) are perhaps the most likely in that
the absorbing column associated with the high temperature component
(presumably due to the supernova) is close to the \rosat\ \pspc\, best
fit value, $\log N_H = 21.7$.  Although the 90\% confidence limits
overlap, these models indicate that the high temperature component
cooled slightly from $kT = 5.1-7.4 \keV$ in 1994 Jan to $kT = 4.1-6.0
\keV$ in 1996 Jan.  In models 2af9X, with the metallicity fixed at
solar, the temperature decrease is slightly more significant since the
90\% confidence limits no longer overlap.  There is also some
indication that the metal abundance may have changed from $0.74-1.4$
solar in 1994 Jan to $0.42-0.90$ solar in 1996 Jan.  Models 2b9X and
2c9X help determine the sensitivity of the two-temperature fits to
errors in the absorbing column associated with the high temperature
component.  Varying the absorbing column between $N_H =
(2.75-7.25)\times 10^{21} \cm^{-2}$, caused the best fit temperature
of the hotter component to vary between $kT = 4.8-10.3 \keV$ in 1994
Jan, and between $kT = 3.99-8.55
\keV$ in 1996 Jan.  Underestimating the absorbing column leads to
large increases in the best fit temperature, but overestimating the
absorbing column leads to only a relatively small decrease in the fit
temperature.  Comparison of models 2a9X and 2ac9X shows that a factor
of 2.4 decrease in the temperature of the cool component decreases the
best fit temperature of the hot component by only about 15\%. 

The measured X-ray fluxes are relatively insensitive to these
uncertainties in the best fit spectrum.  Averaging all the model fits
at the two epochs, we find that the 0.5-2.5 keV flux changed from
about $6.07 \times 10^{-13} \erg \s^{-1} \cm^{-2}$ in 1994 Jan to
about $5.28 \times 10^{-13} \erg \s^{-1} \cm^{-2}$ in 1996 Jan,
corresponding to a decrease of 13.0\%.  The 2-10 keV flux changed from
about $1.62 \times 10^{-12} \erg \s^{-1}
\cm^{-2}$ in 1994 Jan to about $1.17 \times 10^{-13} \erg \s^{-1}
\cm^{-2}$ in 1996 Jan, corresponding to a decrease of 27.8\%.  
If our measured 2-10 keV fluxes are due entirely to the supernova, we
estimate that the unabsorbed supernova luminosity was of $L_X(2-10
\keV) = 2.04 \times 10^{40} \erg \s^{-1}$ in 1994 Jan and $L_X(2-10
\keV) = 1.47 
\times 10^{40} \erg \s^{-1}$ in 1996 Jan, for 10 Mpc distance and $\log
N_H = 21.7$.  

We expect that the 0.5-2.5 keV band contains a strong
constant background due to diffuse emission in the galaxy, but that
the supernova should dominate in the 2-10 keV band, although there may
be some additional background due to unresolved point sources in the
disk.  If the supernova continued to dim $\propto t^{-2}$ between 1994
Jan (age 11.06 yrs) and 1996 Jan (age 13.09 yrs), we would expect the
flux to decrease by $1 - (11.06/13.09)^2 = 28.6\%$, consistent with
the observed decrease in 2.0-10 keV flux.  In the 0.5-2.5 keV band, we
can write the ratio ($r$) of the 1996 Jan flux to the 1994 Jan flux as
\begin{equation}
    r = {s_{96} + g \over s_{94} + g}
\end{equation}
where $s_{94}$ and $s_{96}$ are the fluxes from the supernova at the
two epochs and $g$ is the flux contribution from diffuse emission in NGC 891.
If the supernova is dimming $\propto t^{-2}$, then $s_{96}/s_{94} =
0.714$.  The 1994 Jan ratio of the galaxy flux to the
supernova flux can be written
\begin{equation}
{g \over s_{94}} = \left({1\over f}-1\right)\left({11.06\over t}\right)^2
\end{equation}
where $f \equiv s_t/(s_t + g)$ is the supernova fraction of the total
0.5-2.5 keV emission within a 3\arcmin\, radius of the supernova at age
$t$ years.  Using our \rosat\, data to measure $f$ and assuming that the
supernova is dimming $\propto t^{-2}$ and all other X-ray sources are
constant, we expect the 0.5-2.5 keV flux to decrease by about 10\%,
consistent with the observed decrease of 13.0\% within the errors.
Therefore, the \asca\, flux measurements are consistent with SN1986J
continuing to dim $\propto t^{-2}$ over the energy range 0.5-10 keV
between 1994 Jan and 1996 Jan.

\begin{figure}[h]
\plotfiddle{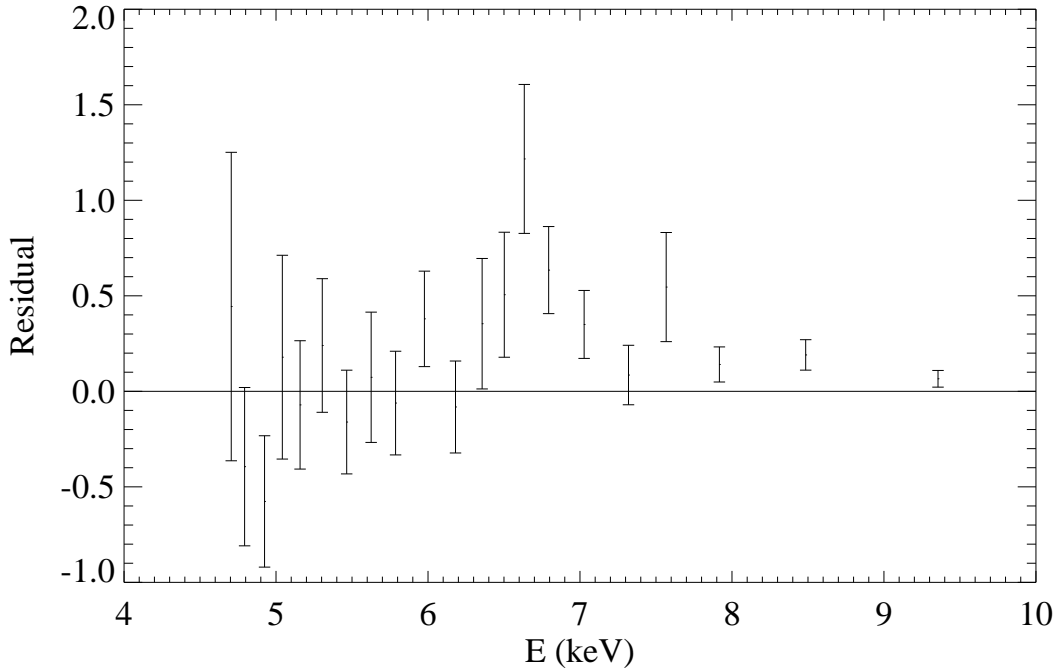}{3.5in}{90}{60}{60}{200}{0}
\caption{Iron line at 6.7 keV shown as residuals after
fitting a 5.37 
keV Bremsstrahlung continuum to the sum of the \siszero\ and \sisone\ spectra
from 1996 Jan.  The residuals are plotted in units of $10^{-3}~{\rm
normalized~counts}~\sec^{-1}~\keV^{-1}$. 
\label{line-residuals}
}
\end{figure} 
 
Because it represents the clearest means of discriminating between the
two proposed models for the X-ray emission, the width of the Fe K line
at 6.7 keV is an important observable parameter.  By fitting a
Gaussian profile to the observed line, we set a 90\% confidence upper
limit of $\sigma=0.17 \keV$ on the line width, for a line centered at
6.79 keV; the best fit line width is $\sigma=0.10
\keV$ (Fig. \ref{line-residuals}). Unfortunately, this upper limit is
not strong enough to discriminate between the model predictions.

\section{Discussion} 

Based on a crude model for the radio light curves of SN 1986J,
Chevalier (1987) estimated a value of $\mdot_{-4}/v_{w1} \sim 1$,
where $\mdot_{-4}$ is the progenitor mass loss rate in units of
$10^{-4} \msun\yr^{-1}$ and $v_{w1}$ is the wind velocity in units of
10 $\kms$, and an explosion date in late 1982.  This model led to the
prediction of X-ray emission from the reverse shock front with a
luminosity $\sim 10^{40} \rm~ergs~s^{-1}$ and a temperature of
$(1-4)\times 10^7$ K.  These predictions were roughly verified by
Bregman \& Pildis (1992).  A more detailed model for the radio
emission by Weiler, Panagia, \& Sramek (1990) led to
$\mdot_{-4}/v_{w1} =2.4$ and an explosion time in late 1982.  However,
this model did involve assumptions about a mixed absorbing/emitting
region and the temperature of the surrounding wind is a factor in
determining the circumstellar density from the radio light curves
(Lundqvist \& Fransson 1988), so this value for $\mdot_{-4}/v_{w1}$ is
uncertain.

CF present details for 2 circumstellar interaction 
models with $\mdot_{-4}/v_{w1} =0.5$ at an age of
10 years.
The red supergiant model has $L=4\times 10^{39} \rm~ergs~s^{-1}$, but a
temperature of only $3\times 10^6$ K because of the steep density
gradient; the shock wave is radiative.
The power law model ($n=8$) has a lower luminosity by a factor of 2,
but the temperature is about $10^7$ K; the shock wave became adiabatic
at an age of 0.1 yr.
In the first case, the luminosity scales roughly as $\mdot/v_w$ and
the second as $(\mdot/v_w)^2$ provided the adiabatic assumption
remains applicable.
The observed temperature suggests that something like the power law
case (i.e. a relatively flat density profile) is more relevant to SN 1986J, but with a higher value of $\mdot/v_w$.
The current X-ray observations point to a value of
$\mdot_{-4}/v_{w1}$ of a few.

A problem with the reverse shock model was raised by Chugai (1993),
who noted that the VLBI observations of Bartel, Rupen, \& Shapiro (1989)
imply an average velocity of 13,000 km s$^{-1}$.
The problem is that an object with typical supernova parameters
expanding into the dense medium discussed above would be decelerated
to considerably lower velocities.
Although Chugai (1993) considers a particular supernova density profile
with $n=8$, it seems that the argument is widely valid.
The 2 models of CF discussed above have peak ejecta velocities of
4300 km s$^{-1}$ and 5000 km s$^{-1}$, respectively, at an age of 10 years.
With the higher circumstellar density inferred for SN 1986J, the
velocities would be even lower.
However, the VLBI image produced by Bartel et al. (1991) in late 1988
shows that the source has a complex structure; it is made up of an
incomplete shell with several protrusions. 
The angular diameter across the peak brightness of the shell is
about 1.6 milliarcsec.
At an age of 6 years, this corresponds to a radial velocity of
6,300 km s$^{-1}$.
The origin of the protrusions is not certain, but the simulations of
Blondin, Lundqvist, \& Chevalier (1996) show that protrusions can develop
if there is an angular density gradient in some direction.
Protrusions would develop most naturally along one axis, which is not the
case for SN 1986J, but the simulations do show the possibility of forming
the required protrusions.

To be more specific, we consider the power law density profile models
discussed in section 2 of CF.
We take $\mdot_{-4}/v_{w1} =2.4$ (Weiler et al. 1990 and the above
discussion of X-ray luminosity) and $n=7$ in order to maximize the
reverse shock gas temperature.
The reference density, $\rho_r$, is defined to be the density at a
velocity of $5,000~\kms$ and age of 1 year, and can be $2\times 10^{-16}
\rm~gm~cm^{-3}$ for a low mass explosion.
Then, at $t=6$ yr, the outer shock velocity is $6,000 \kms$, which
corresponds to an average shock velocity of $7,400 \kms$.
The average velocity in the shell would be somewhat lower than this
and in accord with the VLBI shell velocity.
At $t=10$ yr (late 1992), the outer shock velocity is $5,400~\kms$,
which leads to a post-reverse shock temperature of $1.4\times 10^7$ K.
This temperature is somewhat lower than the observed one,
but the outer shock should have a temperature of $4\times 10^8$ K.
This component should have a lower luminosity than the cool component,
although nonequilibrium effects could give rise to enhanced line
emission.
It is also possible that the supernova density profile is complex and
the shock front is in a particularly flat section.

In this model, the reverse shock makes a transition from radiative
to non-radiative at an age of about 2 years.
At the time of the observations, the X-ray luminosity should tend
to decrease as $t^{-1}$.
The observed decline is steeper than this ($t^{-2}$).
The steeper decline can be produced if the circumstellar density
drops more rapidly than $r^{-2}$ (Fransson, Lundqvist, \& Chevalier 1996).
The very high circumstellar density deduced for SN 1986J suggests
that it may be associated with a short-lived mass loss event prior
to the supernova explosion.
The circumstellar column density outside the
outer shock front is $2.4\times 10^{21} \rm~cm^{-2}$ at an age of 10 yr.
Because of the transition to an adiabatic shock wave and the low
value of $n$, the column density of cool gas behind the reverse
shock front is probably less than this.
The ability of this gas to absorb is further reduced by
hydrodynamic instabilities.
When the above value is compared to the observed values of $N_H$,
it is plausible that the circumstellar medium is a significant
contributor to the total column density to the the supernova.
In this model, the circumstellar $N_H$ declines as $t^{-0.8}$.

The reverse shock model is thus able to account for the general
features of the observed X-ray emission, although it fails to 
reproduce some specific features.
We have not pursued extensions of the model, such as the possibility
of a density gradient dropping more rapidly than $r^{-2}$, because
of the limited data that are available.
In the alternative model of Chugai (1993), the X-ray emission is from
shocked clouds.
There is possible evidence for a clumpy circumstellar medium in that
SN 1986J shows narrow hydrogen emission; the FWHM of the H$\alpha$  line 
corresponds to an expansion velocity of $530\kms$
(Rupen et al. 1987; Leibundgut et al 1991).
However, the observed X-ray temperature requires a shock velocity of
about $2,000\kms$, so that the same clouds cannot be responsible
for the optical and X-ray emission.
The cloud model has many free parameters (the cloud filling factor,
density contrast, size, and the variation of these with radius),
so it is likely that a model fit can be obtained; it is the uniqueness of
the model that is in question.
There is the possibility of observationally distinguishing the models
because the cloud model involves expansion velocities about a factor 3
smaller than those in the reverse shock model.
In both cases, the emission from the receding gas is absorbed by the
supernova, so that the observed emission is primarily blueshifted.
Our
measurement of the Gaussian line width $\sigma$ corresponds to a 90\%
confidence upper limit of $20000 \kms$ FWHM. Unfortunately, this limit
is too weak to distinguish between the two models, but future X-ray
missions, such as {\it AXAF}, may be able to provide an answer.

Although our X-ray light curve so far has only a few points, the light
curve is the least ambiguous piece of evidence because the high
resolution observations give us a high degree of confidence that the
light curve is not seriously contaminated by emission from nearby
X-ray sources.  In contrast, the interpretation of the \asca\, spectra
is somewhat unclear because of the significant level of contamination
from other sources of X-ray emission in the field of view, and in
particular from point sources in the disk and diffuse emission from
the halo of NGC891.  Despite the lack of information on the spectral
characteristics of the contaminating sources, we can estimate the
relative contributions from each source. The diffuse emission from the
X-ray halo is quite soft, with $kT \lesssim 1 \keV$, and contributes
around 40\% of the flux below about 1 keV. While the spectra of the
point sources in the disk are unknown, the change in 2-10 keV flux
observed from 1994 Jan to 1996 Jan is consistent with the supernova
dominating the emission in this band.

X-ray bright supernovae like SN1986J appear to be relatively rare (Houck
\& Bregman 1996). Therefore, this object represents an excellent
opportunity to study the physics of an extreme example of the
interaction of a supernova with a dense circumstellar environment and
we hope to continue monitoring X-ray emission from SN1986J as long as
possible. Despite the relatively rapid decline in X-ray luminosity,
future high spatial resolution observations with \rosat\, and {\it AXAF}
should allow us to track the light curve and to follow changes in the
spectrum for several years to come.
Observations with {\it AXAF} should be particularly useful for providing
high quality spectra and, thus, line widths.

\section{Acknowledgements}

We would like to thank Keith Arnaud, Ken Ebisawa, and the \asca-{\sl GOF} for
their assistance with the \asca data analysis.  Financial support for
this work was provided by NASA grants NAGW-2135, NAGW-4448, 
NAG5-2732, and NAG5-3057.

\end{document}